\documentclass[preprint,a4paper]{aastex631}

\usepackage{amsmath,amssymb,bm}
\usepackage{natbib}
\usepackage{graphicx}
\usepackage{upgreek}

\shorttitle{Flat U-Net}
\shortauthors{Zhu et al.}

\begin{document}

\title{Flat U-Net: An Efficient Ultralightweight Model for Solar Filament Segmentation in Full-disk H\(\alpha\) Images}

\correspondingauthor{GaoFei Zhu}
\email{gfzhu@xhu.edu.cn}

\author[0000-0002-0405-7018]{GaoFei Zhu}
\affiliation{School of Computer and Software Engineering, Xihua University, Chengdu 610039, P. R. China}
\affiliation{National Astronomical Observatories, Chinese Academy of Sciences, Beijing 100101, P. R. China}
\affiliation{State Key Laboratory of Solar Activity and Space Weather, National Space Science Center, Chinese Academy of Sciences, Beijing 100190, P. R. China}

\author{GangHua Lin}
\affiliation{National Astronomical Observatories, Chinese Academy of Sciences, Beijing 100101, P. R. China}
\affiliation{State Key Laboratory of Solar Activity and Space Weather, National Space Science Center, Chinese Academy of Sciences, Beijing 100190, P. R. China}

\author[0000-0003-1675-1995]{Xiao Yang}
\affiliation{National Astronomical Observatories, Chinese Academy of Sciences, Beijing 100101, P. R. China}
\affiliation{State Key Laboratory of Solar Activity and Space Weather, National Space Science Center, Chinese Academy of Sciences, Beijing 100190, P. R. China}

\author{Cheng Zeng}
\affiliation{School of Computer and Software Engineering, Xihua University, Chengdu 610039, P. R. China}

\begin{abstract}

Solar filaments are one of the most prominent features observed on the Sun, and their evolutions are closely related to various solar activities, such as flares and coronal mass ejections. Real-time automated identification of solar filaments is the most effective approach to managing large volumes of data. Existing models of filament identification are characterized by large parameter sizes and high computational costs, which limit their future applications in highly integrated and intelligent ground-based and space-borne observation devices. Consequently, the design of more lightweight models will facilitate the advancement of intelligent observation equipment. In this study, we introduce Flat U-Net, a novel and highly efficient ultralightweight model that incorporates simplified channel attention (SCA) and channel self-attention (CSA) convolutional blocks for the segmentation of solar filaments in full-disk H\(\alpha\) images. Feature information from each network layer is fully extracted to reconstruct interchannel feature representations. Each block effectively optimizes the channel features from the previous layer, significantly reducing parameters. The network architecture presents an elegant flattening, improving its efficiency, and simplifying the overall design. Experimental validation demonstrates that a model composed of pure SCAs achieves a precision of approximately 0.93, with dice similarity coefficient (DSC) and recall rates of 0.76 and 0.64, respectively, significantly outperforming the classical U-Net. Introducing a certain number of CSA blocks improves the DSC and recall rates to 0.82 and 0.74, respectively, which demonstrates a pronounced advantage, particularly concerning model weight size and detection effectiveness. The data set, models, and code are available as open-source resources.

\end{abstract}


~

\section{Introduction} 
\label{sec:intro}

Solar filaments are coronal structures consisting of cold plasma suspended in the hot corona \citep{chen2020some}. Their temperatures typically range from 6000 to 8000 K, aligning with the typical temperature of the solar chromosphere, which makes solar filaments more easily observed in the H\(\alpha\) line. They typically appear as dark, irregular, and elongated structures. Observations indicate that larger, elongated filaments often remain quietly suspended in the corona, gradually evolving until they eventually dissipate. In contrast, smaller filaments may erupt rapidly because of magnetic instabilities, disappear in a short time, and are frequently associated with events such as coronal mass ejections. During the eruption of solar filaments, magnetic field structures on the solar surface and coronal regions undergo significant reorganization, releasing substantial amounts of plasma and high-energy particles that can have a profound impact on space weather. This process may disturb Earth's magnetosphere, posing potential threats to artificial satellites, global navigation systems, and communication networks \citep{camporeale2018machine}. Changes in solar filaments can also provide indirect insights into other solar activities. The formation of solar filaments and whether their material originates from the corona itself are topics of debate in solar physics \citep{parker1953instability,rust1994helical,martin1998conditions,wang1999jetlike,antiochos2000thermal,song2017origin,zhao2017formation,kaneko2017reconnection,wang2018formation,wang2019formation,li2019repeated,chen2020some,jervcic2023dynamic}. Continuous observations produce a substantial amount of H\(\alpha\) data, making automated methods an essential focus for long-term statistical studies. \citet{chen2020some} also mention in their work that automated processing of solar filaments remains a promising research direction for the next decade.

The extraction of solar filaments from H\(\alpha\) observational data using automated methods has always been essential, especially for long-term statistical studies or the analysis of specific physical properties of filaments \citep{hao2015statistical}. Early researchers mainly used various classical image processing methods, focusing on selecting appropriate thresholds to separate filament features from other features in full-disk solar images \citep{gao2002development, shih2003automatic,fuller2005filament, yuan2011automatic,hao2013developing, hao2015statistical}. Primary methods include global threshold, local threshold, and region-growing techniques. Although these approaches offer a clear advantage in computational simplicity, they often lack robustness when applied to more complex scenarios. Traditional machine learning methods, such as support vector machines \citep{labrosse2010automatic} and artificial neural networks \citep{zharkova2005filament}, have also been utilized for detecting solar filaments, focusing primarily on locally cropped images and limb prominences. \cite{delouille2018coronal} proposed an improved version of the Spatial Possibilistic Clustering Algorithm to identify low-intensity regions in extreme-ultraviolet (EUV) images, primarily focusing on coronal holes and filaments, and achieved satisfactory results. \cite{zhu2019solar}, \cite{ahmadzadeh2019toward}, and \cite{salasa2019solar} all proposed using deep learning methods to recognize solar filaments. \cite{salasa2019solar} directly applied the Mask R-CNN network for filament detection, followed by further postprocessing to achieve pixel-level filament segmentation, which yields the expected results. \cite{ahmadzadeh2019toward} adopted the same network model. However, the Mask R-CNN model has a complex structure with a large number of parameters and is primarily designed for object detection, making it relatively coarse for segmentation tasks. Different from the previous two strategies, \cite{zhu2019solar} first attempted using an end-to-end deep learning model, U-Net, to extract solar filaments from H\(\alpha\) full-disk solar images. By incorporating random dropout layers, their model reduces the risk of overfitting caused by excessive parameters, and is characterized by an elegant architecture and simplicity of implementation. Additionally, their approach effectively mitigates issues such as excessive noise that is often generated by traditional methods. Since then, many researchers have continued to advance along the direction of deep learning to improve its accuracy and architecture. \citet{liu2021solar} further enhanced the accuracy of solar filament segmentation by modifying the padding operation, expanding the pathway, and incorporating the ASPP module, thus effectively minimizing the occurrence of low-level noise. \cite{guo2022solar} achieved impressive results by combining various feature extraction modules (ResNet-C, ResNet-D, and ResNet v2) into a new backbone for CondInst \citep{tian2020conditional}, which serves as a feature extractor for H\(\alpha\) images. \citet{diercke2024universal} trained two models, YOLOv5 \citep{jocher2020ultralytics} and U-Net, to first detect bounding boxes for solar filaments and then perform pixel-level segmentation. \citet{jiang2024automated} achieved the expected segmentation results by integrating an attention mechanism into the U2-Net architecture. In addition, \cite{mackovjak2021scss} proposed the SCSS-Net model, which is based on U-Net, for detecting coronal holes in EUV images, showcasing the strong potential of the U-Net model in various astronomical tasks.

Upon reviewing recent studies employing deep learning for solar filament segmentation in H\(\alpha\) full-disk images, we note that most researchers have aimed to make their models as intricate or expansive as possible, often by incorporating additional modules or integrating multiple models. However, the increasing size and complexity of these models may hinder practical application. Particularly for the development of highly integrated deep neural networks for intelligent ground-based or space-based instruments (including convolutional neural networks (CNNs)), models are primarily built by stacking a variety of linear and nonlinear functions to process and learn from existing data. The back-propagation algorithm \citep{rumelhart1986learning} adjusts the function parameters to facilitate the learning process, repeating this until the loss function converges. This highlights the strong correlation between the problem scale and the model scale. Larger models are often used to tackle smaller-scale problems, which can result in many ineffective parameters. For general tasks, larger and more complex models are often preferred. In particular, the feature extraction modules of these networks typically increase both depth and width, but this can also introduce issues like vanishing gradients. To address these issues, additional residual connections are typically added to preserve as much of the features as possible \citep{he2016deep}. However, this approach can also increase the likelihood of introducing many ineffective parameters just to achieve a certain level of effective weights. To counteract this, commonly used models often integrate dropout layers or carry out operations like pruning and quantization on existing models. In the specific task of extracting solar filaments from H\(\alpha\) full-disk images, although the random variations in solar filaments may increase the problem scale, the relatively uniform and stable background of solar images actually reduces the overall complexity of the task. Based on this, it is more worthwhile to explore finding a suitable model for the specific task.

Inspired by the self-attention concept from \cite{vaswani2017attention}, we propose an efficient and ultralightweight network model, Flat U-Net \citep{zhu2025flatunet}, for filament segmentation in full-disk H\(\alpha\) solar images. Concretely, our main contributions are as follows: (1) Channel self-attention convolutional block (CSA-ConvBlock) is designed to reallocate weights among feature map channels, enhancing the effectiveness of interchannel weights, while simplified channel attention convolutional block (SCA-ConvBlock) is also proposed to address hardware limitations during training. (2) In the new U-Net model, which is composed of (SCA)CSA-ConvBlocks, the number of channels in the encoder and decoder is no longer doubled but becomes flattened, which is the origin of the name ``Flat U-Net.'' This strategy significantly reduces the model parameters to around 1M while maintaining an elegant structure and still achieving effective solar filament segmentation results. (3) A new solar filament data set has been established, allowing full-disk H\(\alpha\) images to retain disturbances such as limb darkening and moderate cloud cover. Our proposed method can segment solar filaments using only a small number of model parameters, without requiring any modifications to the original FITS image content. 

The structure of this paper is organized as follows: The proposed model architecture and (SCA)CSA-ConvBlock are described in detail in Section~\ref{sec:methods}. Section~\ref{sec:exp} is primarily focused on the data, implementation details, and experimental results. The discussion and conclusions are presented in Section~\ref{sec:dis_con}.

\section{Methods} 
\label{sec:methods}

\subsection{Architecture Overview} 
\label{subsec:architecture}

\begin{figure}[!htbp]
\centering
\plotone{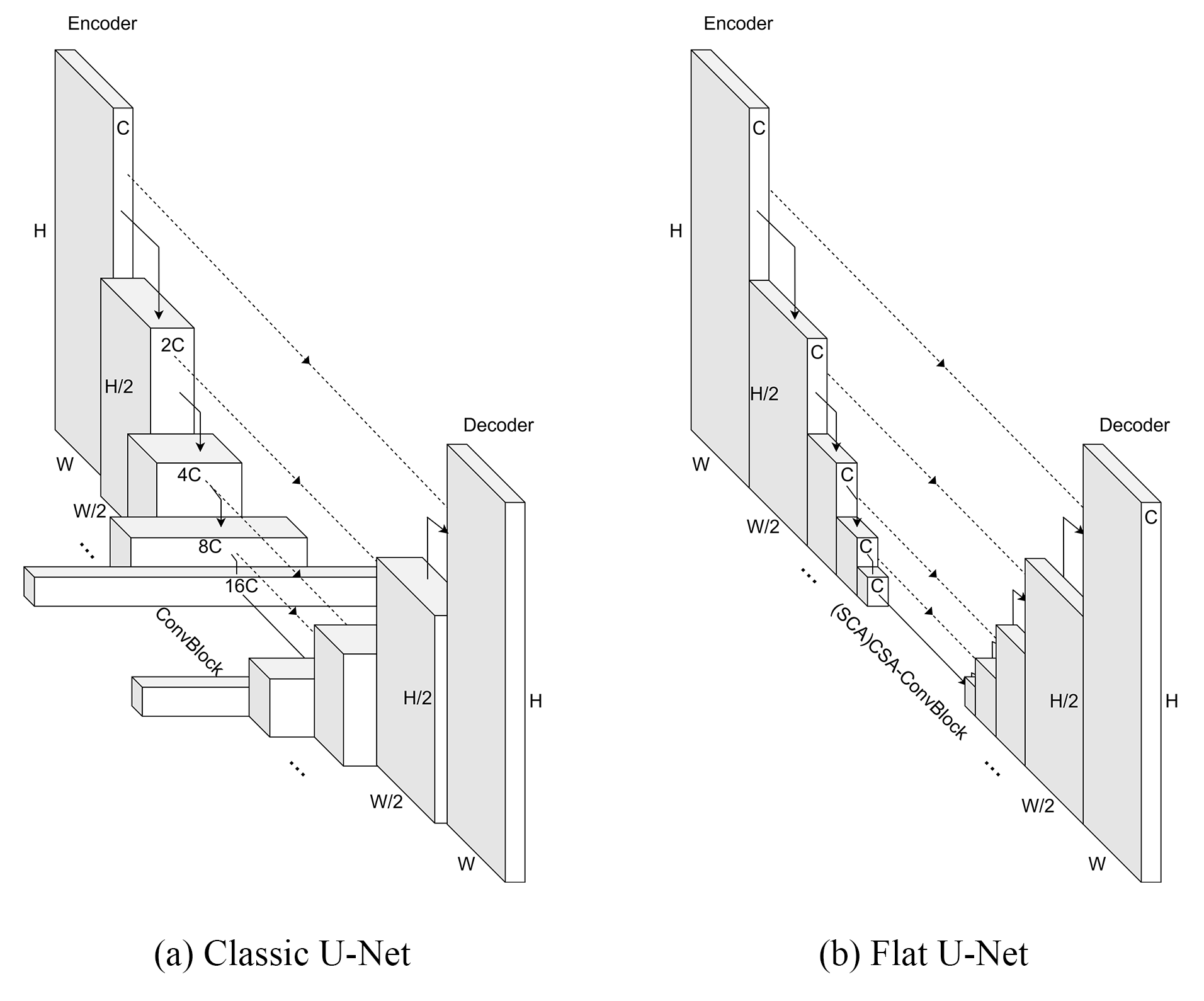}
\caption{The proposed Flat U-Net architecture is compared with the classical U-Net. The overall network structure of Flat U-Net, which is composed of the proposed (SCA)CSA-ConvBlock modules, optimizes interchannel weight efficiency and minimizes redundant parameters. The design presents a flattened architecture.}
\label{fig:architecturefig}
\end{figure}

In this study, we propose a new network model, Flat U-Net (Figure~\ref{fig:architecturefig}(b)), in which flattening serves as a prominent feature compared to the classical U-Net model (Figure~\ref{fig:architecturefig}(a)). This architecture comprises an encoder and a decoder, which are employed to extract and recover feature information from full-disk H\(\alpha\) solar images, respectively. The input to the network is full-disk H\(\alpha\) solar images in FITS format, which require only min--max normalization without the removal of limb darkening and unevenness correction (e.g., cloud cover). Initially, a classical convolution module is employed to expand the single-channel solar image to \(\mathit{C}\) channels, and the number of channels is maintained consistently at \(\mathit{C}\) throughout the network. Subsequently, the input is fed into the encoder, which comprises four layers, constructed from (SCA)CSA-ConvBlock modules. The encoder primarily extracts features from the solar images while progressively reducing the spatial resolution. In the encoder, (SCA)CSA-ConvBlock is a crucial component that models dependencies among feature channels, effectively improving the model performance in feature extraction. This mechanism reallocates channel weights in the feature maps using self-attention, improving the model's capacity to capture interchannel dependencies. Compared to classical CNN architectures, Flat U-Net retains essential features while minimizing the number of parameters, thereby achieving a lightweight design. The decoder gradually restores the features extracted by the encoder, returning the spatial resolution to match that of the input image. The skip connections are identical to those found in the classical U-Net. 

A notable feature of Flat U-Net architecture is its lightweight design. Each encoder and decoder block incorporates an (SCA)CSA-ConvBlock that optimizes the channel distribution of feature maps to reduce the number of parameters. This approach avoids the redundancy mechanism in the classical U-Net structure, where the number of channel features increases as the spatial features of the feature maps decrease, leading to an increase in redundant weight parameters. In the task of segmenting solar filaments in full-disk H\(\alpha\) solar images, the reallocation of feature channels through the (SCA)CSA-ConvBlock module enables the model to locate and segment the fine filaments. This efficient design not only reduces the model's storage requirements but also significantly decreases inference time, thereby enhancing real-time processing capabilities.

\subsection{Channel Self-attention Convolutional Block} 
\label{subsec:csa}

\begin{figure}[!htbp]
\centering
\plotone{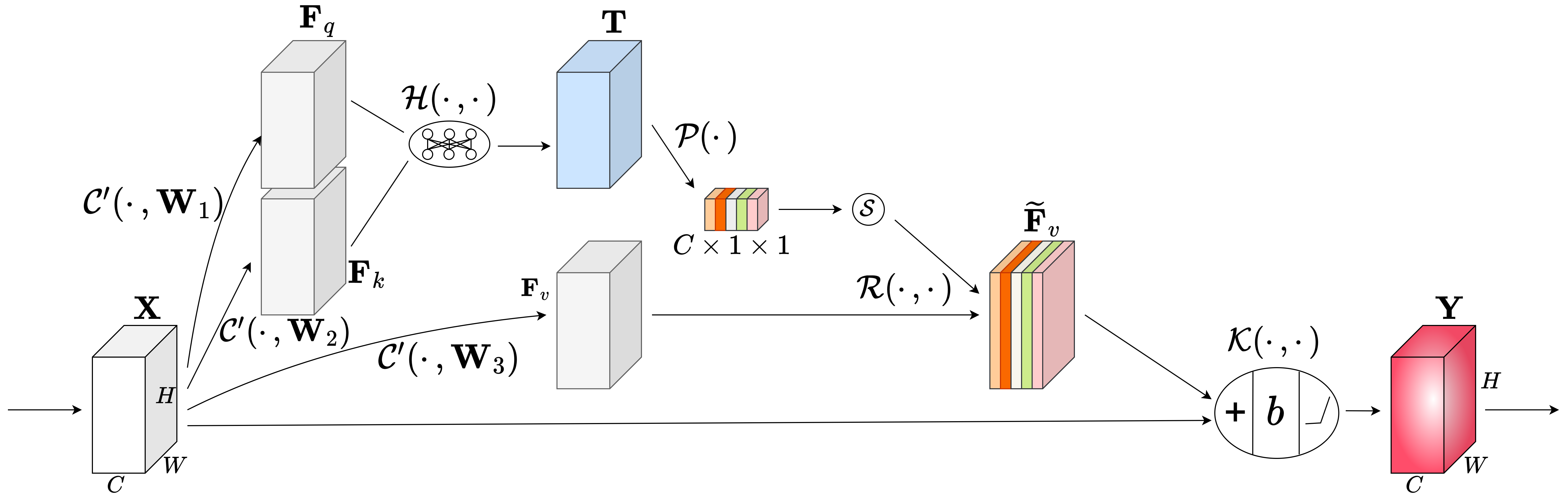}
\caption{CSA-ConvBlock.}
\label{fig:csablockfig}
\end{figure}

A preliminary preprocessing step is necessary for the original input single-channel full-disk H\(\alpha\) solar image \(\bm{X}'\). The single channel must be expanded to multiple channels, which can be expressed as:
\[
\bm{X} = \mathcal{C}(\mathrm{Norm}(\bm{X}')),
\]
where \(\bm{X}' \in \mathbb{R}^{1 \times H \times W}\) and \(\bm{X} \in \mathbb{R}^{C \times H \times W}\). \(\mathrm{Norm}\) represents the min--max normalization operation and \(\mathcal{C}\) denotes the convolution operation. The transformation of the CSA-ConvBlock (Figure~\ref{fig:csablockfig}) that constitutes the entire network can be defined as:
\[
\mathcal{F}: \bm{X} \to \bm{Y}; \quad \bm{X}, \bm{Y} \in \mathbb{R}^{C \times H \times W},
\] 
which implements a series of operations designed to enhance feature extraction. 

In CNNs, the application of pooling operations decreases the ability of filters to extract effective information from feature maps. To address this, we leverage the concept of self-attention mechanisms to enhance the effectiveness of the parameters within the block. The main focus is to apply this strategy to multichannel feature maps, reallocating the weights of each channel to enable the filters to learn more effective parameters. First, an unbiased convolution operation denoted as \(\mathcal{C}'\) is applied simultaneously to the input block \(\bm{X}\), resulting in the feature maps: query \(\bm{F}_q\), key \(\bm{F}_k\), and value \(\bm{F}_v\). This process is expressed mathematically as:
\[
\bm{F}_q = \mathcal{C}'(\bm{X}, \bm{W}_1), \quad \bm{F}_k = \mathcal{C}'(\bm{X}, \bm{W}_2), \quad \bm{F}_v = \mathcal{C}'(\bm{X}, \bm{W}_3),
\]
where \(\bm{X} \in \mathbb{R}^{C \times H \times W}\) represents the input tensor, \(\bm{W}\) is the convolution kernel, and \(\bm{F}_q, \bm{F}_k, \bm{F}_v \in \mathbb{R}^{C \times H \times W}\) denote the resulting feature maps. The effectiveness of filter parameters is enhanced by increasing the correlation of feature maps among channels. The similarity tensor \(\bm{T}\) is generated by applying \(\bm{F}_q\) and \(\bm{F}_k\) through broadcasting, where the computation for the \(c\)th channel is as follows:
\[
{{\bm{T}}^{(c)}} = \mathcal{H}({\bm{F}}_q^{(c)},{{\bm{F}}_k}) = {\bm{F}}_q^{(c)}\cdot{{\bm{F}}_k} = \frac{{\mathop \sum \limits_{s = 1}^C \left( {{\bm{F}}_q^{(c)}\cdot{{({\bm{F}}_k^{(s)})}^{\text{T}}}} \right)}}
{{\sqrt {H[W]} }},
\]
where \(\bm{F}_q^{(c)}\) represents the feature map of the \(c\)th channel in \(\bm{F}_q\). The normalization factor, \(\sqrt{H}\), is set to balance the scale differences arising from features of different dimensions, thus stabilizing the gradients. \(H\) denotes the height of the feature map, and in the current work, \(H\) and \(W\) are equal. Next, global average pooling is applied to the spatial information, compressing the similarity tensor \(\bm{T}\) into channel-wise scores, which can be expressed as:
\[
\bm{s}_c = \mathcal{P}(\bm{T}^{(c)}) = \frac{1}
{{H \times W}}\sum\limits_{i = 1}^H {\sum\limits_{j = 1}^W {{\bm{T}^{(c)}}(i,j)}}, 
\]
where \(\bm{s} \in \mathbb{R}^{C}\) is a global feature descriptor along the channel dimension. \(\widetilde{ \bm{s}} = [{\widetilde{s}_1} ,{\widetilde{s}_2 }, \ldots,  \widetilde{s}_c] \in {\Delta ^{c-1}}\) is obtained by applying the softmax function \(\mathcal{S}\) to the vector \(\bm{s}\). Next, We reconstruct the channel-wise interaction representation of the feature maps \(\bm{F}_v\) using the scores \(\widetilde{\bm{s}}\):
\[
\widetilde{\bm{F}}_v  = \mathcal{R}(\widetilde{\bm{s}},\bm{F}_v) = \widetilde{\bm{s}}_i \cdot \bm{F}_v^{(i)} = [\widetilde{s}_1 \cdot \bm{F}_v^{(1)}, \widetilde{s}_2 \cdot \bm{F}_v^{(2)}, \ldots ,\widetilde{s}_c \cdot \bm{F}_v^{(c)}],
\]
where \(\widetilde{\bm{F}}_v \in \mathbb{R}^{C \times H \times W}\) matches the dimensions of tensor \(\bm{X}\). Another key point that cannot be ignored is the residual connection between \(\widetilde{\bm{F}}_v\) and the input \(\bm{X}\) at the final stage of the block:
\[
\bm{Y} = \mathcal{K}(\widetilde{\bm{F}}_v, \bm{X}) =  g(b(\widetilde{\bm{F}}_v + \bm{X})),
\]
where \(+\) denotes the addition of tensors with the same dimensions, \(b\) is used for normalizing the residual result \citep{ioffe2015batch}, and \(g\) refers to the ReLu function \citep{nair2010rectified}. Through the aforementioned methods, establishing connections among different convolutional operations within each block is more effective for feature extraction than merely stacking them. The network is composed of CSA-ConvBlocks, in which the dimensions of the input \(\bm{X}\) and output \(\bm{Y}\) remain consistent, and the feature information of the individual channels in the input \(\bm{X}\) is reallocated.

\subsection{Simplified Alternative} 
\label{subsec:sa}

Due to the extensive matrix operations involved in computing interchannel correlations, particularly under conditions of limited hardware resources such as GPU memory, we have proposed a simplified alternative. Replacing CSA with SCA significantly reduces GPU memory usage during the training process. The computation of the \(c\)th channel feature is changed to:
\[
\bm{T}^{(c)} = \mathcal{H}(\bm{F}_q^{(c)},\bm{F}_k^{(c)}) = \frac{\bm{F}_q^{(c)} \cdot (\bm{F}_k^{(c)})^{\text{T}}}{\sqrt{H[W]}},
\]
where \({\bm{F}}_q^{(c)}\) and \({\bm{F}}_k^{(c)}\) represent the \(c\)th channel features in \({\bm{F}}_q\) and \({\bm{F}}_k\), respectively. This effectively handles slightly lower hardware environments without changing the number of parameters.

\section{Experiments} 
\label{sec:exp}

\subsection{Data} 
\label{sec:data}

The validation experiment was conducted to verify the effectiveness of the proposed method without limb-darkening removal and unevenness corrections. The original data were downloaded from the Big Bear Solar Observatory (BBSO) website in FITS file format. They consist solely of raw full-disk solar images centered in the field of view. The ground truth maps are binary mask images obtained using traditional methods. However, these methods have certain limitations (such as redundant noise and excessive preprocessing), leading to inconsistent quality in the generated results. Therefore, only the results that meet the quality criteria are selected to serve as data set samples of ground truth. Even though labeling errors are present, they do not impede the model proposed in this paper from efficiently learning the features of solar filaments. Approximately 1000 images from 2022 to 2023 were selected as training and validation sets, including FITS images and their corresponding PNG masks. Additionally, approximately 100 FITS files with ground truth were selected from 2024, along with around 150 FITS files without ground truth from 2021, to form the test set. This choice was primarily due to the potential inaccuracy of the ground truth, which could lead to misunderstandings when used as a comparative evaluation benchmark. We will also select a portion of images with cloud cover as a second test set to evaluate the model's effectiveness, employing manual visual inspection for evaluation.

\subsection{Implementation Details} 
\label{sec:impl}

The deployment environment of the proposed Flat U-Net is Python 3.8, PyTorch 1.12, and CUDA 11.3. Other combinations of compatible versions can also be applied to this method. During training, standard data augmentation techniques, such as flipping and rotation, are applied to increase the diversity of the data. The input size ($B$, $C$, $H$, $W$) of the network is related to the memory capacity of the GPUs and the batch size \(B\) during training. The raw data downloaded from BBSO have dimensions of \(2048 \times 2048\), which poses a considerable challenge for a single Nvidia 4080 GPU with 16 GB of memory. Although \(1024 \times 1024\) can also be trained normally, it requires setting the batch size to 2, which will take considerably more time. Taking all factors into account, we have opted to set the input size to \(512 \times 512\), and the batch size is 9. Notably, to obtain effective lightweight model weights, our method involves a significant amount of tensor multiplication operations during the computation process. As a result, the memory requirements during training will be somewhat higher. To accommodate the current experimental conditions, we chose a combination of SCA and CSA as the training strategy, where SCA serves as the backbone network for feature extraction, and CSA acts as the bottleneck. The network can also be adapted to specific scenarios, consisting entirely of either CSA or SCA, with CSA generally yielding better performance than SCA. The other configuration is set as follows: the optimizer is Adam \citep{kingma2014adam}, with an initial learning rate of \(10^{-3}\) and dynamic learning rate adjustment, and the loss function is binary cross entropy.

\subsection{Results} 
\label{sec:res}

As mentioned previously, the proposed model is flattened, but what would be the appropriate thickness (channels) for this network model in the context of solar filament segmentation? We separately validate the convergence of the loss during training and the inference performance in the test set under different numbers of channels \(C \in \{ n \in \mathbb{Z}^+ \mid n > 1 \}\). Under the existing hardware conditions, we selected \(C \in [8, 9, 12, 16, 32, 64, 128, 256]\) for testing. For inference evaluation, we use metrics such as dice similarity coefficient (DSC), precision, recall, and F1-score for relative comparison, which are represented, respectively, as follows: 
\[
\text{DSC} = \frac{2 \times \text{TP}}{2 \times \text{TP} + \text{FP} + \text{FN}},
\]
\[
\text{Precision} = \frac{\text{TP}}{\text{TP} + \text{FP}},
\]
\[
\text{Recall} = \frac{\text{TP}}{\text{TP} + \text{FN}},
\]
\[
\text{F1-score} = \frac{2 \times \text{Precision} \times \text{Recall}}{\text{Precision} + \text{Recall}},
\]
where TP, TN, FP, and FN represent true positive, true negative, false positive, and false negative, respectively. This work involves a small-target binary classification task, and inevitably includes instances of erroneous annotations. In the training set, there are instances with features that are not filaments labeled as filaments, and as well cases with features that are unlabeled filaments. Through testing, we found that the filament features learned by the model can sometimes be more accurate than the labeled information (as will be presented later). Therefore, each metric has only a relative reference value.

\begin{figure}[!htbp]
\centering
\plotone{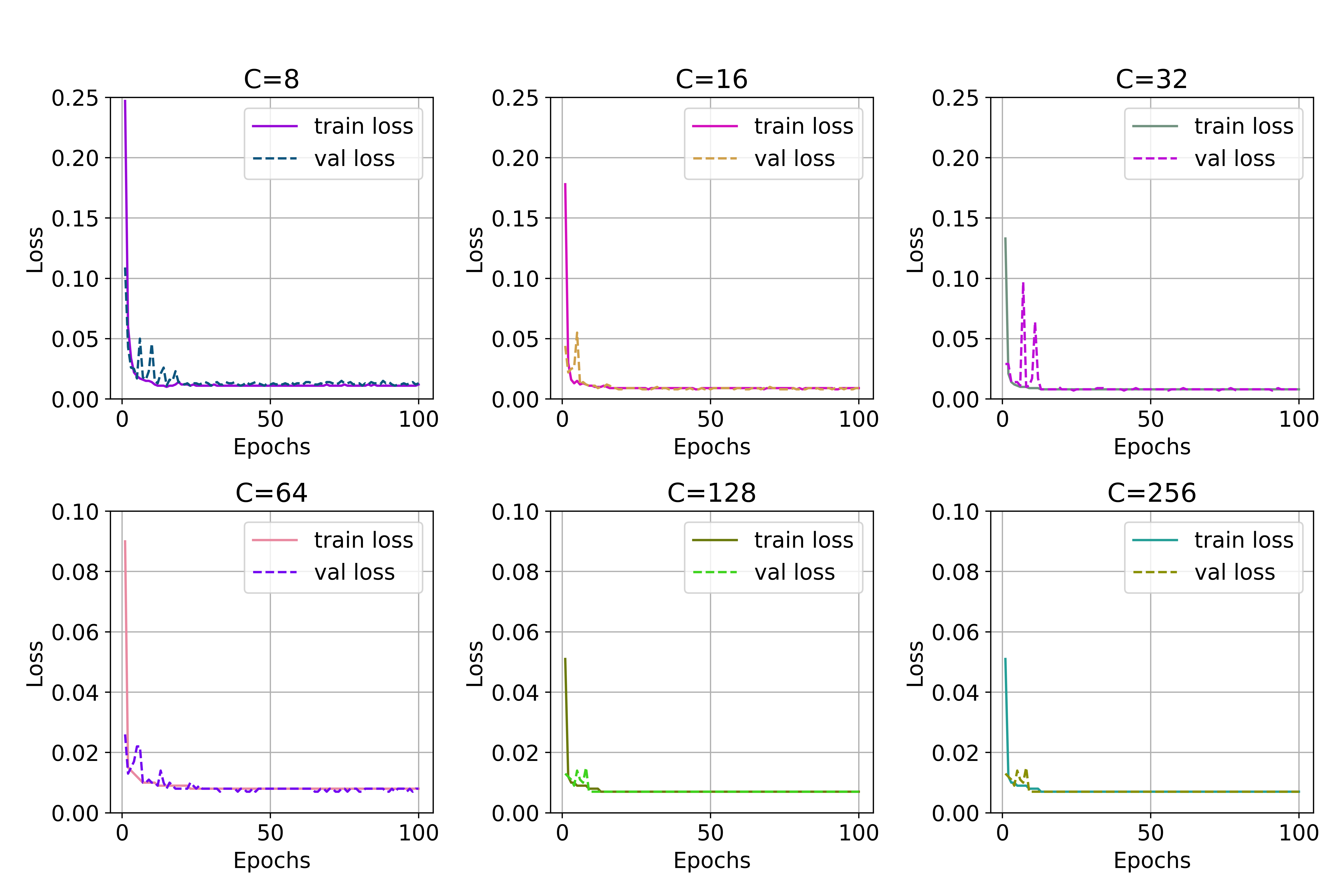}
\caption{The training process of the proposed model with different umbers of channels.}
\label{fig:lossfig}
\end{figure}

As shown in Figure~\ref{fig:lossfig}, when \(C=8\), the training process converges; however, the inference metrics in Table~\ref{tab:t1} indicate that the model fails to learn any meaningful features. This suggests that the model's parameter count and complexity are insufficient for the task. In contrast, when \(C=16\), the metrics improve significantly, with precision reaching 0.93, indicating that the model begins to demonstrate learning capability when \(C\) is between 8 and 16. Further experiments reveal that the model starts acquiring learning ability around \(C=9\) (refer to Table~\ref{tab:t1}). This indicates that the proposed model can achieve the task of identifying solar filaments in full-disk H\(\alpha\) solar images with only 0.08 MB (float32) of memory for its parameters. From other metrics, it can be seen that when \(C<16 \), there are slight fluctuations in the results due to the random selection of training data and data augmentation operations. When the model has the ability to learn, increasing the number of parameters can improve the metrics to some extent.

\begin{table}
\centering
\caption{Performance with Varying Numbers of Channels}
\label{tab:t1}
\begin{tabular}{l|ccccccc}
\toprule
Models & Params (\(10^6\)) & Model Size (MB) & FPS\footnote{Frames Per Second.} (float32) & DSC & Precision & Recall & F1-score \\
\hline
Flat U-Net(8) & -- & -- & -- & 0.0 & 0.0 & 0.0 & 0.0 \\
Flat U-Net(9) & 0.02 & 0.08 & 450 & 0.72 & 0.91 & 0.60 & 0.72 \\
Flat U-Net(12) & 0.04 & 0.14 & 430 & 0.76 & 0.90 & 0.64 & 0.75 \\
Flat U-Net(16) & 0.07 & 0.25 & 430 & 0.73 & 0.93 & 0.60 & 0.73 \\
\hline
\end{tabular}
\end{table}

Figure~\ref{fig:lossfig} also reveals that as the number of channels increases, the model demonstrates a good convergence capability. Table~\ref{tab:t2} indicates that the overall improvement in metrics tends to slow down. From 16 to 256 channels, the precision is maintained at around 0.93, with the former having a model size of just 0.25 MB (float32), while at $C=256$, the model size sharply increases to 62.78 MB. Improvement in other metrics is also very limited. This indicates that an excessive number of parameters no longer significantly contributes to improving the metrics, leading to an abundance of redundant parameters. Table~\ref{tab:t2} indicates that when \(C=32\), the model shows good segmentation capability, with the size of the model being less than 1 MB. Subsequently, as the number of parameters increases, the performance improvement gradually slows down or even fluctuates (due to the randomness of the data). Compared with the classical U-Net, as the overall number of channels decreases, most metrics experience a substantial decline, with only precision remaining stable. Furthermore, when using the [8, 16, 32, 64, 128] channel configuration, the model loses its ability to learn for the current task. While the classical U-Net achieves a precision of approximately 0.98, its other metrics are generally lower than those of Flat U-Net, indicating a higher miss rate for solar filament detection compared to Flat U-Net. Table~\ref{tab:t2} further demonstrates that, for both the classical U-Net and Flat U-Net, once the number of model parameters reaches a certain level, further increases lead to minimal or no significant improvements in performance. In contrast, if the number of parameters falls below a critical line, the performance metrics will experience a sharp decline, ultimately causing the model to lose its learning ability. In general, Flat U-Net possesses learning capability with only 0.25 MB of parameters. Furthermore, it achieves relatively stable recognition results with less than 1 MB of parameters (\(C=32\)), highlighting its efficiency, lightweight, and relative accuracy.

\begin{table}[!htbp]
\centering
\caption{Comparison with the Classical U-Net}
\begin{tabular}{l|ccccccc}
\toprule
Models & Params (\(10^6\)) & Model Size (MB) & FPS (float32) & DSC & Precision & Recall & F1-score \\
\hline
U-Net [64, ..., 1024] & 28.95 & 110.45 & 330 & 0.69 & 0.98 & 0.53 & 0.68 \\
U-Net [32, ..., 512] & 7.24 & 27.63 & 340 & 0.69 & 0.98 & 0.54 & 0.68 \\
U-Net [16, ..., 256] & 1.81 & 6.92 & 410 & 0.59 & 0.98 & 0.43 & 0.59 \\
U-Net [8, ..., 128] & -- & -- & -- & 0.0 & 0.0 & 0.0 & 0.0 \\
Flat U-Net(8) & -- & -- & -- & 0.0 & 0.0 & 0.0 & 0.0 \\
Flat U-Net(16) & 0.07 & 0.25 & 430 & 0.73 & 0.93 & 0.60 & 0.73 \\
Flat U-Net(32) & 0.26 & 0.98 & 380 & 0.79 & 0.93 & 0.69 & 0.79 \\
Flat U-Net(64) & 1.03 & 3.93 & 250 & 0.77 & 0.92 & 0.67 & 0.77 \\
Flat U-Net(128) & 4.12 & 15.7 & 220 & 0.81 & 0.93 & 0.72 & 0.81 \\
Flat U-Net(256) & 16.5 & 62.78 & 200 & 0.79 & 0.94 & 0.68 & 0.79 \\
\hline
\end{tabular}
\label{tab:t2}
\end{table}

\begin{figure}[!htbp]
\centering
\includegraphics[width=0.7\textwidth]{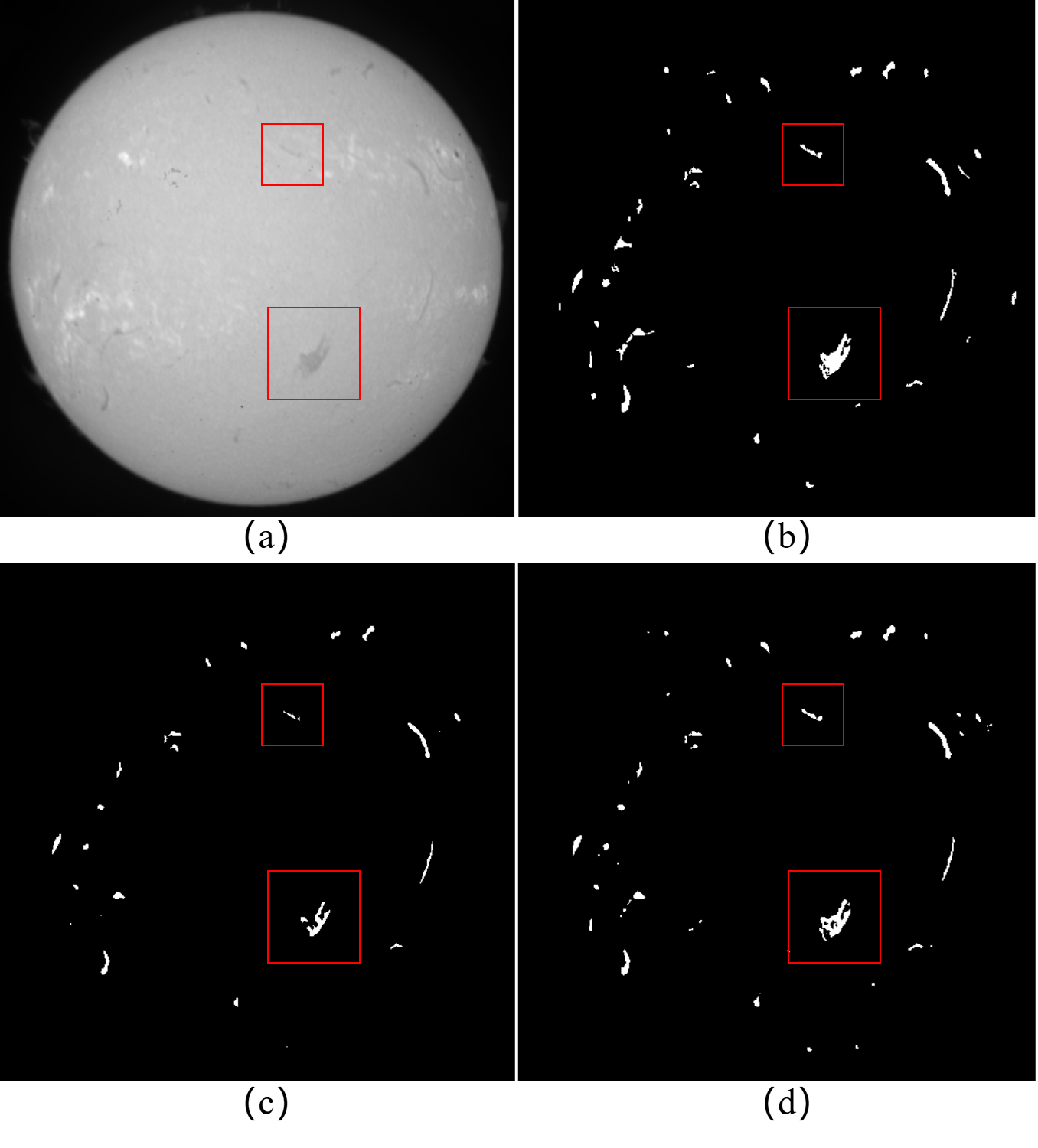}
\caption{An example of visual comparison across different methods. (a) Full-disk H\(\alpha\) solar image (without limb-darkening correction) from BBSO, captured on 2024 March 4, at 17:44:55 UT. (b) Ground truth map. (c) Segmentation result from U-Net with channel configuration [64, ..., 1024]. (d) Segmentation result from lat U-Net with channel configuration \(C=32\).}
\label{fig:cmpfig}
\end{figure}

\begin{figure}[!htbp]
\centering
\includegraphics[width=0.7\textwidth]{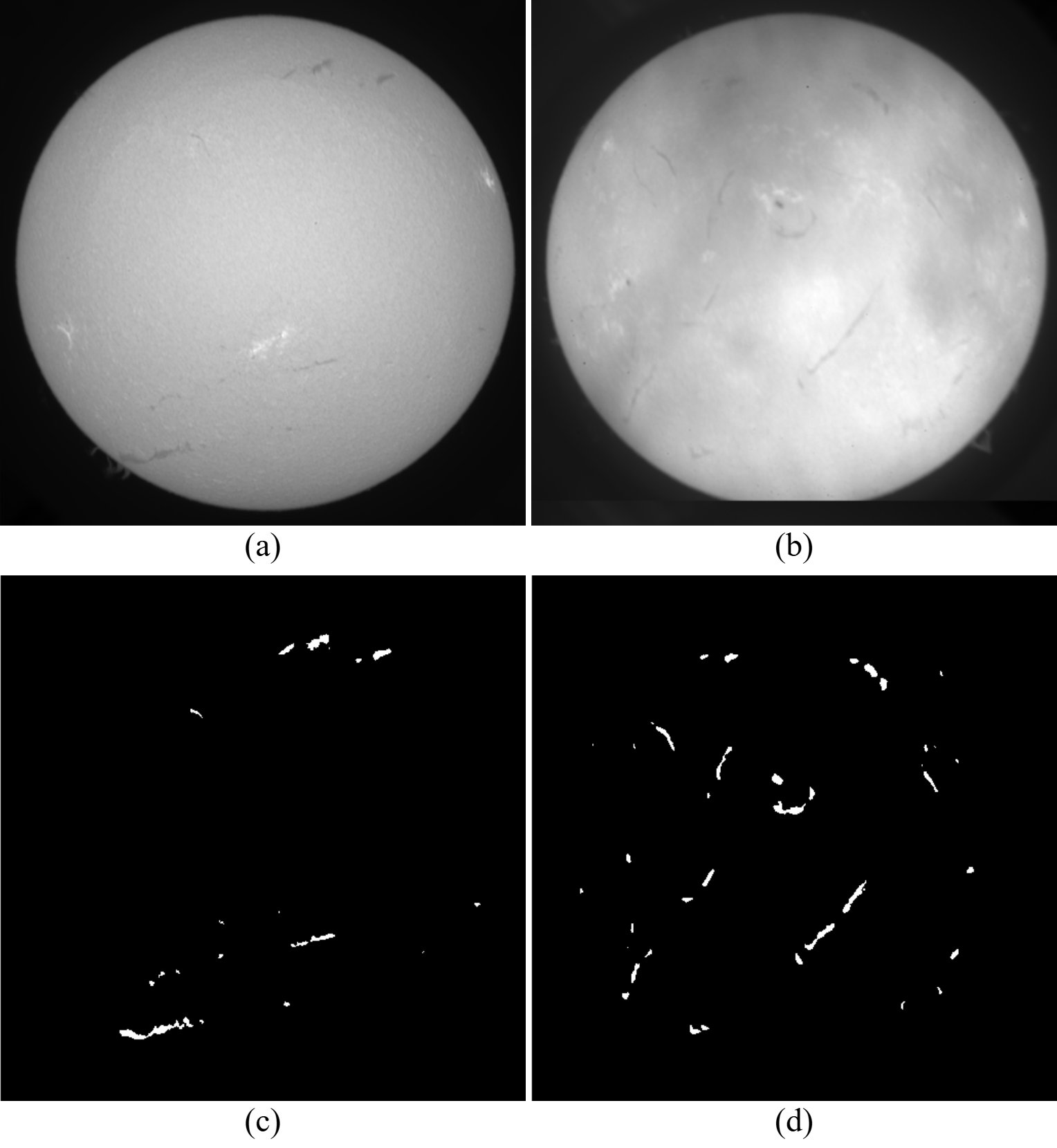}
\caption{Examples of segmentation results without ground truth using the available model trained by \(C=32\). (a) Clear full-disk H\(\alpha\) solar image. (b) Full-disk H\(\alpha\) solar image contaminated by clouds. (c) Segmentation results from panel (a). (d) Segmentation result from panel (b).}
\label{fig:nogdfig}
\end{figure}

Figure~\ref{fig:cmpfig} shows an example of segmentation results from different methods. Although U-Net can identify solar filaments in full-disk H\(\alpha\) solar images, the results exhibit certain instances of missed detections, including both small and large filaments. This issue primarily arises from factors such as limb darkening, which obscures the characteristics of the filaments, as well as the inherent inefficiencies of U-Net's parameters. Flat U-Net achieves results superior to those of the previous model, even with an extremely small number of parameters. This is due to the efficient capture of solar filament features across different layers of the model. In addition, we also applied the proposed method to unlabeled data, including clear images and those contaminated by clouds. Figure~\ref{fig:nogdfig} illustrates a segmentation example for each case. It can be seen that even though the training set contains few labeled data with cloud contamination, the trained lightweight model can still effectively recognize solar filament features. The data set, models, and code are available in the National Astronomical Data Center at doi: \href{https://doi.org/10.5281/zenodo.14610155}{10.12149/101545}.

Table~\ref{tab:t3} presents a comparison of Flat U-Net (\(C=32\)) under different block configurations. The model is composed of four convolutional blocks (each with four in the encoder and decoder, respectively) and a bottleneck. As the number of CSA blocks increases, metrics such as DSC and recall improve significantly, while precision shows a slight decline. This suggests that more true filament regions are being correctly identified, albeit with a slight increase in background regions mistakenly classified as filament regions. It is worth noting that when the entire network is composed of SCA, the various metrics achieve relatively good results. However, upon adding the first CSA to the bottleneck, recall increases significantly from 0.6402 to 0.6948, and DSC rises from 0.7581 to 0.7943. This demonstrates that CSA provides a clear advantage over SCA in the bottleneck configuration. In other words, a more efficient reconstruction of interchannel feature information can mitigate the loss of spatial features in the bottleneck. Increasing the number of CSA blocks inevitably results in relatively higher computational complexity compared to SCA. Therefore, the trade-off between segmentation precision and computational cost becomes evident as the number of CSA blocks increases. While the improved DSC and recall highlight the model's enhanced ability to capture filament structures, the slight decrease in precision suggests an increased tendency to include nonfilament regions. This trade-off highlights the importance of carefully selecting the number of CSA blocks, particularly in contexts with stringent computational constraints or high precision requirements.

\begin{table}[!htbp]
\centering
\caption{Comparison of Different Configurations of Flat U-Net}
\begin{tabular}{l|cccccc}
\toprule
Block Configuration & Params (\(10^6\)) & Model Size (MB) & DSC & Precision & Recall & F1-score \\
\hline
SCA(5) and CSA(0) & 0.26 & 0.98  & 0.7581 & 0.9396 & 0.6402 & 0.7581 \\
SCA(4) and CSA(1) & 0.26 & 0.98  & 0.7943 & 0.9333 & 0.6948 & 0.7933 \\
SCA(3) and CSA(2) & 0.26 & 0.98  & 0.8112 & 0.9273 & 0.7248 & 0.8112 \\
SCA(2) and CSA(3) & 0.26 & 0.98  & 0.8175 & 0.9214 & 0.7385 & 0.8175 \\
SCA(1) and CSA(4) & 0.26 & 0.98  & 0.8185 & 0.9211 & 0.7413 & 0.8185 \\
\hline
\end{tabular}
\label{tab:t3}
\end{table}

\section{Discussion and Conclusion} 
\label{sec:dis_con}

With the continuous advancements and upgrades in astronomical observation hardware, vast amounts of observational data have been generated, making automated feature extraction one of the promising research fields. A lightweight design not only meets the need for efficient processing of large-scale data but also provides technological support for advancing highly integrated and intelligent systems in ground-based and space-based instruments. In network models, pooling strategies are commonly used to reduce computational complexity. However, this often results in the loss of spatial information, which necessitates an increase in the number of channels to mitigate this loss and enhance feature representation. We propose a practical ultralightweight network architecture, named Flat U-Net, which efficiently leverages the spatial information within each channel by reconstructing interchannel features. The architecture consists of (SCA)CSA-ConvBlocks, which are designed to extract filament features from full-disk H\(\alpha\) solar images. It can directly extract solar filaments from raw H\(\alpha\) images even for data contaminated by clouds.  

Experimental results demonstrate that the proposed method requires only 0.25 MB for the model to achieve learning capability, with a precision exceeding 0.9. When the number of channels \(C=32\), the model's overall performance reaches a relatively stable level. Compared to the classical U-Net, the proposed method shows advantages in metrics such as DSC, recall, and F1-score, with only a slight reduction in precision, indicating superior performance and promising prospects. The significantly higher recall compared to the classical U-Net demonstrates that the proposed method is more effective at identifying solar filament regions marked by masks. This improvement is mainly attributed to the fact that the features in each layer are no longer obtained through simple stacking but are reconstructed through interchannel feature maps, making feature extraction in each convolutional block more efficient.

According to the comparison experiments in Table~\ref{tab:t3}, we find that the entire network composed of SCA achieves relatively good precision. Additionally, as the number of CSA blocks increases, DSC, recall, and F1-score show significant improvements. Based on the previous analysis, we recommend configuring at least the bottleneck of the model with CSA, while other configurations can be freely combined.

Finally, the proposed method offers the following advantages: (1) Efficiency and lightweight design are the primary characteristics of the proposed network, allowing it to learn more profound filament features with an extremely small number of parameters. There is no need for additional trick operations, and the comprehensive performance of the segmentation, with network weights under 1 MB, shows an advantage over the classical U-Net, making it highly suitable for deployment in practical scenarios. (2) The proposed (SCA)CSA-ConvBlocks optimize the feature representations transmitted between network layers by reconstructing interchannel relationships. This approach mitigates the disadvantage of needing to increase the number of channels to compensate for the loss of spatial feature information caused by pooling operations within the network. (3) This approach overcomes the issue of suboptimal performance when directly applying the classical U-Net model to raw full-disk H\(\alpha\) solar images, eliminating the need for limb-darkening removal and unevenness correction. Moreover, the model framework presented in this paper is applicable to data from other solar features, and it also shows significant potential for similar scenarios in other fields.

\section*{Acknowledgments}
The authors greatly appreciate the constructive comments provided by the referees. Our research is supported by the Talent Introduction Program of Xihua University (Z242123). It is also funded by National Key R\&D Program of China (2022YFF0503800), National Natural Science Foundation of China (12173049), and projects from National Astronomical Observatories, CAS (H242522 and H242079). The authors thank BBSO for providing the full-disk H\(\alpha\) data.


\bibliography{zgf24}{}
\bibliographystyle{aasjournal}



\end{document}